\documentclass{article}

\usepackage{booktabs}
\usepackage{array}
\usepackage{enumitem}
\usepackage{float}
\usepackage{placeins}
\usepackage{tikz}
\usetikzlibrary{arrows.meta,positioning}
\usepackage[T1]{fontenc}
\usepackage[utf8]{inputenc}
\usepackage{lmodern}

\usepackage{placeins}

\usepackage{adjustbox}
\usepackage{caption}
\usepackage{subcaption}
\usepackage{amsmath}

\usepackage{graphicx} 

\title{Decision-Aware Trust Signal Alignment for SOC Alert Triage}
\author{
\begin{tabular}{cc}
\textbf{Israt Jahan Chowdhury} &
\textbf{Md Abu Yousuf Tanvir} \\
Ontario Tech University, Canada &
Ontario Tech University, Canada
\end{tabular}
}
\date{}

\begin{document}

\maketitle

\begin{abstract}

Machine learning-based detection systems are being used more and more in Security Operations Centers (SOCs) to sift through security alerts in vast amounts. In practice, probabilistic outputs or confidence scores are frequently revealed by these systems, but these signals are frequently miscalibrated and unable to be interpreted when pressure is applied in practice. The results of prior qualitative and survey based research on SOC practice indicate that alarm quality and alert overload are related to the burden and difficulty of making decisions by the analyst, particularly when the outputs of the tools are noisy or otherwise hard to act on in a consistent manner [3],[4]. The main weakness is that the concept of confidence is often stated without the explicit correspondence to asymmetric decision costs whereby false alarms are far less harmful than an attack missed.

The paper presents a decision conscious trust signal alignment framework of SOC alert triage. Rather than changing the detection models, the framework maps (i) the calibrated confidence, (ii) lightweight uncertainty cues and (iii) cost sensitive decision threshold to a coherent decision support layer [6]. Calibration is based on common post-hoc strategies demonstrated to enhance the consistency of probabilistic forecasts [5], and uncertainty cues offer conservative protection in cases the model predictions are uncertain [6].

To measure model-agnostic effects in our evaluation of the approach on the UNSW-NB15 intrusion detection benchmark [1],[2] we use two different classifiers Logistic Regression and Random Forest. Simulation results show that misaligned confidence displays significantly increase false negatives, while decision-aligned trust signals reduce cost-weighted loss by orders of magnitude. We also give out a human-in-the-loop study protocol so that future analyst assessment of SOC such as triage tasks can be made.
\end{abstract}


\section{Introduction}

\subsection{Background}
In the current Security Operations Centers (SOCs) are founded on automated systems of detection to observe infrastructures of enormous complexity to identify malicious activity in large volumes. The application of machine learning in the intrusion detection has evolved to be a common practice in the sense that machine learning can model high dimensional network telemetry on as well as adapt to various attack patterns [7],[8]. The models of detection used within a working environment never work independently: this creates a steady flow of alerts, which have to be filtered, examined and raised by a human analyst.Unlike traditional IDS research, this work does not propose a new detection model, but instead focuses on how existing model outputs are interpreted and acted upon in SOC environments.

Some of the never-ending ones include the alert fatigue which is a mixture of a high number of alerts, the frequency of false alarms, and time pressure that causes degradation of the attentions of the analysts and consistency of decision making [4]. Also important according to the empirical descriptions of the SOC work is that quality of the alarm has significant impact on the actions of the analysts and the workload, and that the quality of the signal can also determine the quality of the triage, and of the incident detected [3],[1]. In that regard, the probability of false negative (missed attacks) does not proportionally correlate with the false positive (additional investigation) and therefore SOC triage is actually cost sensitive.
\subsection{Problem Statement}
To enable most SOC tools, the model outputs are produced in the form of predicted labels, rankings or confidence scores. Although these signs may be viewed as a sign of credibility, they are not usually oriented towards making cost-conscious decisions. When confidence is not in place, the high and low scores can be over- and under-trust by the analysts, and this can be dangerous. In a larger sense, there is no such thing as actionable trust, that is, confidence: a probability may be statistically, but operationally, inaccurate unless it is linked to the effects of escalation versus closure.

\subsection{Research Gap}
The majority of the IDS studies focus on enhancing the accuracy of models at the model level, detection rate, or AUC [7],[8]. Simultaneously, there has also been the development of trustworthy-AI research that has produced calibration techniques and uncertainty estimation techniques [5],[6], and explainable AI techniques that provide model-agnostic ways of interpreting predictions [9],[10]. But the directions never seem to be united in a decision centric SOC setting. Specifically, trust signals (confidence, calibration, uncertainty) are scarcely studied as modifying the triage decision and there is little empirical work on the existing systems that match the presented trust displays with actual cost models of missed attacks and false alarms. This drives a paradigm that considers trust as a decision support construct, as opposed to a crude model deliverable.

\subsection{Contributions}
The following contributions are made in this paper:
\begin{itemize}
    \item We present a decision mindful trust signal correspondence system of SOC alert triage that explicitly indicates trust signals to the operational decision expenses.
    \item We show that false negative rates can be increased even in the case of misaligned confidence displays, when underlying detection models are kept constant.
    \item We propose a trust alignment mechanism that is cost-sensitive and uses post-hoc calibration with uncertainty-sensitive safeguards to make decisions on escalation.
    \item We are offering model-agnostic empirical validation on both Logistic Regression and Random Forest classifiers on the benchmark of UNSW-NB15.
    \item To facilitate future research on the use of analysts or trained proxies, we design a human-in-the-loop evaluation protocol that is appropriate to SOC style triage.
\end{itemize}

\section{Related Work}
\subsection{ML-Based Intrusion Detection}
The intrusion detection systems (IDS) based on machine learning have been widely researched as the method of malicious behavior recognition within large scale networks [29]. The evaluation of detection methods under controlled environments has been made possible by public benchmark datasets like KDD99, NSL-KDD, CIC-IDS and UNSW-NB15 [1],[2]. Of them, UNSW-NB15 is considered to be more realistic with its up to date attack scenarios and more features representation.

Many different models of machine learning and deep learning have been suggested to be used as an IDS, among which there are logistic regression, support vector machines, random forests, convolutional neural networks, and recurrent architectures [7],[8]. The main focus of this literature has been on the enhancement of detection accuracy, recall and area under curve. Although it is true that these advances have greatly enhanced predictive performance. They have mostly assumed that detection is an independent classification task, and little thought has been given to the way in which model outputs are used by human operators in operational triage [11],[29].

\subsection{Trustworthy and Explainable AI}
In parallel with the development of better detection performances, the area of trustworthy AI has investigated the means of enhancing the dependability and explainability of machine learning systems. Probability calibration methods attempt to achieve the property that when predicting scores on confidence scales, predicted scores represent actual probabilities of accuracy, and attempt to fix systematic overconfidence or under confidence in current models [5],[18],[27].These approaches are becoming of importance to decision support systems in which probabilistic results are used to guide the human judgment [21],[28].

Predictive feature explanations have been used extensively as explainable AI (XAI) methods such as LIME and SHAP [9],[10],[19],[25]. The techniques enhance transparency by bringing out the influential features and decision rationales. Nevertheless, explanations are not the only reason why one should have appropriate trust. In a previous study, it has been demonstrated that users can misunderstand explanations. By developing excessive trust in seemingly interpretable models even when the predictions are not reliable [24]. Therefore, explanation can not be confused with calibrated or decision aligned trust.
\subsection{Human and AI Collaboration in Security}
The research on the interaction between security analysts and automated alerting systems in actual SOC setups has been studied recently. The qualitative studies indicate that analysts tend to get bombarded with alert in large quantities, erratic quality of alerts and inadequate contextual clues. Those are all leading factors to alert fatigue and impaired decision-making [3],[4],[18]. These papers focus on the fact that the behavior of the analysts is influenced by the accuracy of the models, as well as the presentation of the alerts and the confidence signals [22].

Human in the loop security systems have been suggested to integrate automated and human judgment, especially in high impact decisions. Though these systems do not ignore the role of an analyst in monitoring, it is often assumed that it is enough to provide an output of the models or explanations [17],[18],[19]. Empirical studies of the effect that various designs of trust signals have on the decisions by analysts. To escalate those are scant, particularly in asymmetric cost set-ups [26].
\subsection{Decision-Focused and Cost-Sensitive Learning}

The cost sensitive learning deals with the case where various classification errors are penalized differently. False negatives are usually much more expensive than false positives in security applications.For this reason,threshold and loss-sensitive optimization strategies are encouraged [14],[15]. Previous research has suggested decision theoretic methods of varying classification thresholds or learning goals in relation to a cost of errors.
In all these developments, cost-sensitive reasoning is hardly manifested in the trust indicators of the operational systems as seen by the users. Cost conscious thresholds are typically implicit in models, as opposed to being revealed as decision support interface. Consequently, analysts can be given statistically valid confidence scores that are not aligned to operational risk.

However, these approaches stop at model correctness and do not address how trust signals influence cost-sensitive human decisions.

Overall, the current literature has developed intrusion detection accuracy, model sensitivity, and cost-sensitive learning separately. Nonetheless, no single framework has been developed that incorporates the trust signals, decision costs, and human-in-the-loop triage in the SOC environments. Specifically, existing systems do not have systems to match the asymmetric costs of escalation decisions with displays of confidence and uncertainty. This void prompts the decision conscious trust signal alignment framework presented in the given paper.
\section{Proposed Framework}

This part presents a decision conscious trust signal alignment framework of SOC alert triage. The essence is to make sure that trust indicators made to analysts like confidence. Uncertainty are brought to the front and directly adjusted to the asymmetric costs of security choices, as opposed to the agents being considered inert model outputs.
\subsection{Problem Formulation}
The SOC alert triage is a binary decision problem to us. For each alert
the analyst needs to make a choice about:

\begin{equation}
d_i \in \{\text{Escalate}, \text{Close}\}
\end{equation}

where the escalation begins an additional investigation and dismissal of the alert by closure.
\subsubsection{Cost Asymmetry}

The asymmetry of costs of decision errors occurs operationally.The closing of a malicious alert by a false negative (FN) can cause a drastic effect and a false positive (FP) closing of a benign alert will mainly increase the amount of work that an analyst performs. This inconsistency is in line with existing cost-sensitive learning formulations in security and risk-sensitive decision systems [11].

This is formalized by a model of costs:

\begin{equation}
C_{\mathrm{FN}} \gg C_{\mathrm{FP}}
\end{equation}

\subsubsection{Trust Signal Mismatch}

Contemporary detection systems usually provide a predicted label and a score \begin{equation}
p(y = 1 \mid x_i)
\end{equation} of the label. But, this confidence is usually, Miscalibrated, and no longer tied to the Costs of Operational Decision. Consequently, analysts can either over-put their trust or under-put their trust on alerts resulting in disproportionate misjudgment of the results of their decisions. This discrepancy is why there is an incentive to institute trust signals that are clearly aimed at facilitating decision-making in the context of asymmetric cost.
\subsection{Trust Signals}

The suggested framework obtains a collection of model-agnostic lightweight trust signals on each alert.
\subsubsection{Confidence}

The base confidence signal is the posterior probability, from equation-3, generated by the detection model. Raw confidence, which is commonly applied, does not provide enough information on which to base reliable decision-making in SOC situations.

\subsubsection{Calibration}
A systematic overconfidence or underconfidence can be corrected by post-hoc calibration of the raw probabilities with standard methods [5].Calibration generates a corrected probability ($p_{\mathrm{cal}}$) which is a more accurate estimate of the possibility of malicious activity.

The need of calibration is due to the fact that the decision thresholds formed based on uncalibrated scores might not be aligned with the actual risk especially when the output of the models is directly interpreted by a human analyst [5].

\subsubsection{Uncertainty}

Besides confidence, discrete uncertainty signal is obtained in the framework based on calibrated probabilities. Alerts that have probabilities closer to the decision boundary will be classified as high uncertainty and alerts that are distant to the decision boundary will be classified as low uncertainty.

This uncertainty signal gives a protective hedge in case of doubt, which is an expression of the fact that uncertain notifications should not be eliminated just due to slight differences in confidence [6].
The framework can be used alongside post-hoc methods of explanation like LIME or SHAP [9],[10] and feature level explanations can be presented. Nonetheless, explanations are considered to be the additional context rather than the main trust indicators, acknowledging the fact that interpretability is not the guarantee of adequate trust. 
\subsection{Trust Alignment Aware of decisions}
The key contribution of this writing is the correspondence between trust cues and the cost of decision.
\subsubsection{Cost-Sensitive Thresholding}

Given the cost model $C_{\mathrm{FN}}$ and $C_{\mathrm{FP}}$, we obtain a decision-conconscious threshold:

Let $C_{\mathrm{FN}}$ and $C_{\mathrm{FP}}$ denote the operational costs associated with false negatives and false positives, respectively. Under this cost model, we derive a decision-aware classification threshold given by:

\begin{equation}
t^{*} = \frac{C_{\mathrm{FP}}}{C_{\mathrm{FP}} + C_{\mathrm{FN}}}
\label{eq:cost_sensitive_threshold}
\end{equation}

This threshold directly represents the asymmetric cost of errors and substitutes the usual fixed thresholds (e.g. 0.5) which are operationally risk-neutral.
Trust Signal into Decision Mapping.The threshold is associated with the underpricing of the anticipated decision cost with a 0-1 loss weighted against asymmetric operational risk.

In case of escalation it is advisable to raise an alert in case: $p_{\mathrm{cal}} \geq t^{*}$

This mapping makes sure that the confidence values are seen in the context of the cost of making a decision but not as the absolute values of rightness.
Uncertainty Safety Escalation.

To reduce further the false negativity, the framework implements a safety rule, which is to escalate an alert with high uncertainty even when the calibrated confidence of this alert is a bit less than the threshold [11]. This is a conservative approach, which reflects the SOC practice where vague alerts should be investigated more.
\subsection{Interface Conditions}
In order to examine the impact of alignment of trust signals, we are going to distinguish three interface conditions which are differentiated in the presentation and usage of trust signals.
\subsubsection{Baseline Interface - C0}
The base or original condition only shows the forecasted class label. Decisions are implicitly associated with a set threshold and they do not reveal confidence, uncertainty, and cost data.

\subsubsection{Mismatched Trust Interface - C1}
The poor match situation exhibits uncalibrated confidence that is not in line. It is assumed that analysts excessively rely on confidence values and use a more rigid escalation rate, as is common in the operation when the interpretation of the confidence is not applied in the context of the decision.
\subsubsection{Proposed Trust Interface - C2 (Aligned)}
The offered condition would have a calibrated confidence, uncertainty made explicit and decision recommendation made based on the cost sensitive threshold. The trust signals are directly brought in line with operational risk, which steers the analysts to less risky escalation decisions.
\subsection{System Overview}

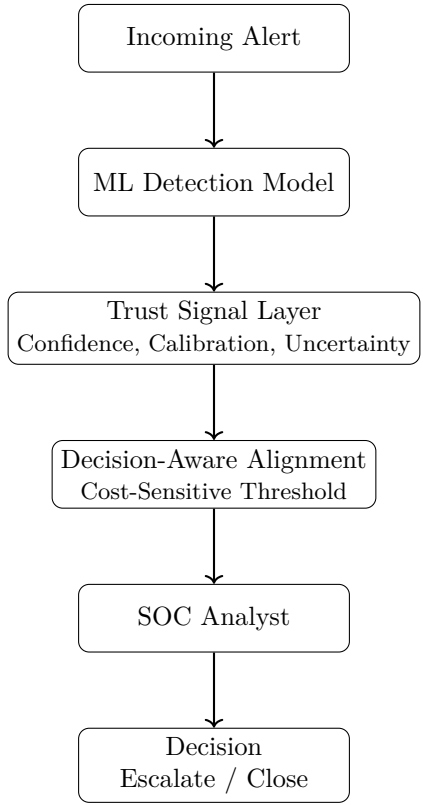
\begin{figure}[t]
\centering
\begin{tikzpicture}[
    box/.style={
        rectangle,
        draw,
        rounded corners,
        align=center,
        minimum width=3.6cm,
        minimum height=0.9cm
    },
    arrow/.style={
        ->,
        thick
    }
]

\node[box] (input) {Incoming Alert};
\node[box, below=of input] (model) {ML Detection Model};
\node[box, below=of model] (trust) {Trust Signal Layer\\
\small Confidence, Calibration, Uncertainty};
\node[box, below=of trust] (decision) {Decision-Aware Alignment\\
\small Cost-Sensitive Threshold};
\node[box, below=of decision] (analyst) {SOC Analyst};
\node[box, below=of analyst] (output) {Decision\\
Escalate / Close};

\draw[arrow] (input) -- (model);
\draw[arrow] (model) -- (trust);
\draw[arrow] (trust) -- (decision);
\draw[arrow] (decision) -- (analyst);
\draw[arrow] (analyst) -- (output);

\end{tikzpicture}
\caption{Overview of the proposed decision-aware trust signal alignment framework for SOC alert triage.}
\label{fig:framework}
\end{figure}

The suggested decision conscious trust signal alignment model is exemplified in Figure 1. Through calibration and uncertainty estimation, model predictions are then converted to trust signals. Such signals are then matched with a cost sensitive decision policy and followed up by presenting it to the analyst. The framework has the advantage of explicitly decoupling detection and decision support such that trust signals are understood in the sense of operational risk.

\section{Experimental Evaluation}
This section empirically investigates the idea of whether the alignment of trust signals to the operational decision costs can enhance SOC alert triage results. In contrast to Section 3 that exposes the framework in an abstract way, the section is entirely concerned about the way the framework is instantiated, implemented and tested in a real experimental context.
\subsection{Dataset and Alert Stream Construction}
This will be done through the construction of dataset and alert stream.

UNSW-NB15 is a collection of intrusion detection experiments that are performed using the UNSW-NB15 intrusion detection data set that was constructed to capture current attack patterns, as well as to mirror real background traffic [1],[2],[12]. A record is associated with a network flow and is considered as a single SOC alert, which has a binary ground-truth label of benign or malicious activity.

We adhere to the initial benchmark division and rely on the prearranged test part as a simulation of an alert stream. In order to minimize the amount of confounding variables and to be able to compare the results of different models, only numerical characteristics are kept, which is a similar approach to prior empirical research on UNSW-NB15 [7],[12]. Alerts are handled separately and are similar to typical SOC triage processes where analysts process alerts one at a time with no model retraining or feedback.
\subsection{Prediction Outputs and Detection Models}
Two classifiers are employed to investigate the resilience of the suggested approach in two sets of models:

\begin{itemize}
    \item Among the linear probabilistic bases, the Logistic Regression (LR).
    \item Random Forest (RF) which is a non-linear ensemble classifier.
\end{itemize}

The two models are also trained using the same set of features and generate posterior probabilities.
$p(y = 1 \mid x_i)$ for each alert. Notably, the probabilities are considered as definite input to the evaluation: at the time of the test, no retraining, threshold adjustment, and loss reweighting are done. The effect of alignment of trust signals and decision logic is isolated in this design, and not the enhancement of the detection capability.
\subsection{Trust Signal Construction}

\begin{itemize}
    \item Calibration: Since probabilistic outputs are often poorly calibrated with real-world Machine Learning systems, we use post-hoc calibration to convert raw probabilities into calibrated probabilities [5],[13],[14].

    Concretely in our pipeline:
    \begin{itemize}
        \item A sigmoid-shaped calibration of LR outputs is used (Platt-style calibration).
        \item The isotonic regression (a non-parametric monotonic calibrator) is used to calibrate RF outputs.
    \end{itemize}
\end{itemize}
\begin{itemize}
    \item Uncertainty: Demonstrating using fixed bands has given us an uncertainty category:
    \begin{itemize}
    \item \textbf{High uncertainty:} $p_{\mathrm{cal}} \in [0.45, 0.55]$
    \item \textbf{Medium uncertainty:} $p_{\mathrm{cal}} \in [0.35, 0.45) \cup (0.55, 0.65]$
    \item \textbf{Low uncertainty:} otherwise
\end{itemize}

\end{itemize}

These categories of uncertainty are deliberately coarse decision cues on interfaces, as opposed to probabilistic guarantees, which, again, are reflective of the workflows that face the analyst.

\begin{figure}[t]
\centering
\begin{subfigure}{0.48\linewidth}
  \centering
\includegraphics[width=0.9\linewidth]{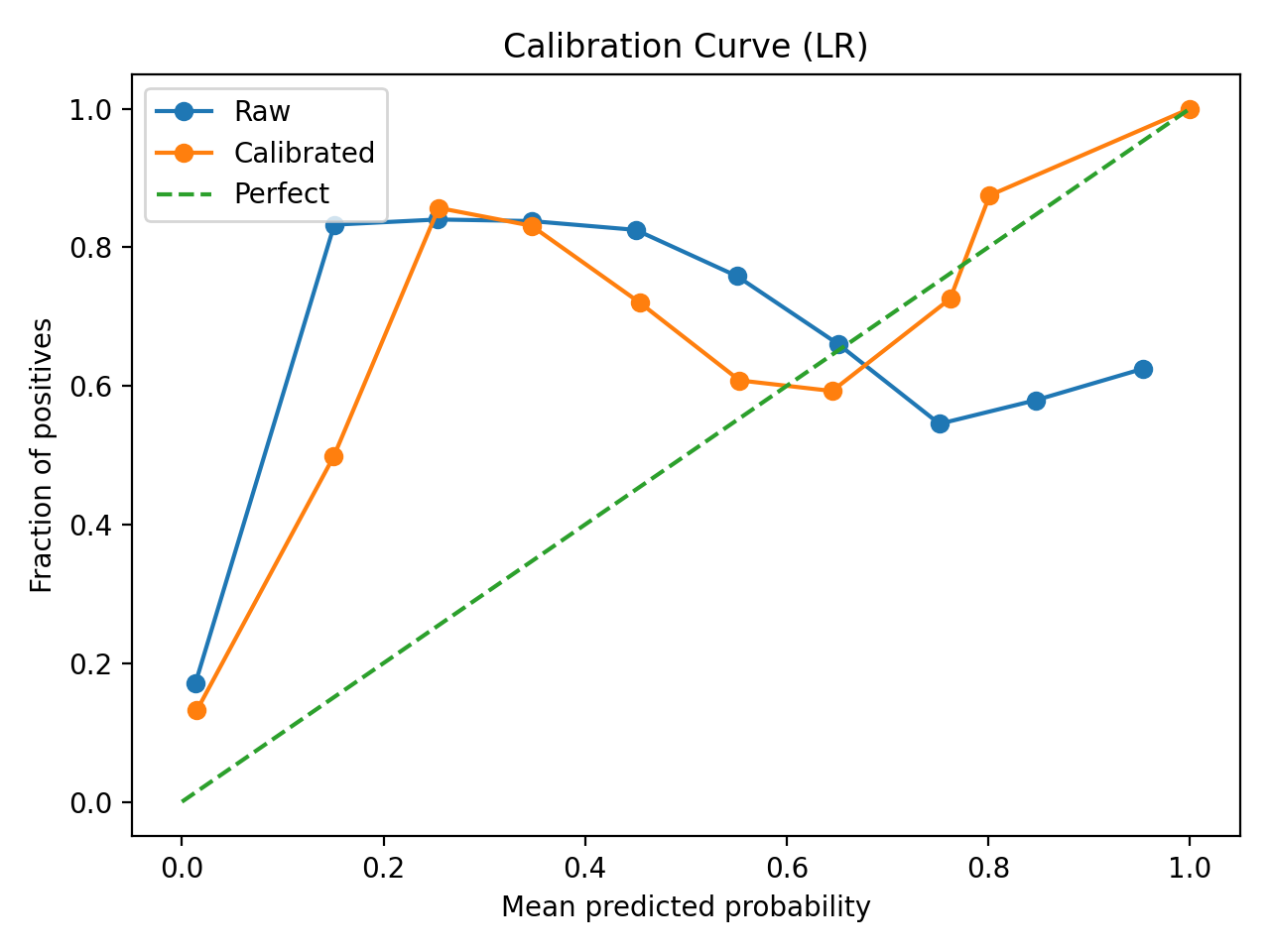}
  \caption{Logistic Regression (LR).}
  \label{fig:cal_lr}
\end{subfigure}
\hfill
\begin{subfigure}{0.48\linewidth}
  \centering
 \includegraphics[width=0.9\linewidth]{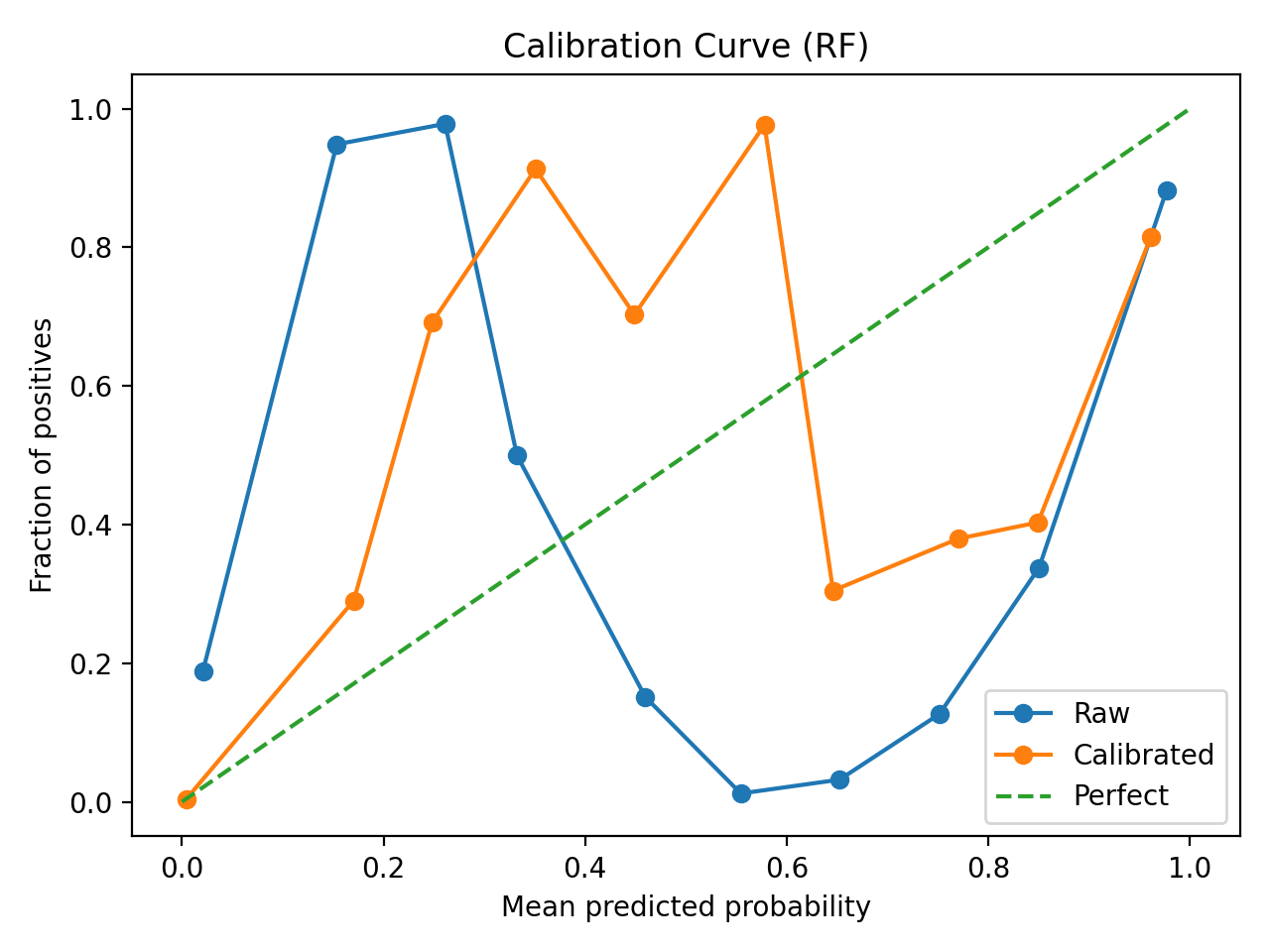}
  \caption{Random Forest (RF).}
  \label{fig:cal_rf}
\end{subfigure}
\caption{Reliability diagrams comparing raw and calibrated probability estimates. The dashed diagonal indicates perfect calibration.}
\label{fig:calibration}
\end{figure}

We apply post-hoc probability calibration so that model confidence can be interpreted as an approximate likelihood of malicious activity. 
Figure~\ref{fig:calibration} reports reliability diagrams for LR and RF, comparing raw probability outputs to calibrated probabilities against the ideal diagonal.
Calibrated probabilities $p^{cal}$ are used as the confidence signal in the aligned trust condition (C2), while the misaligned condition (C1) exposes raw probabilities [16].

The bands are deliberately low-level in order to be easily readable as interface indicators and to provide conservative decision-making when the model is almost indifferent [6],[9],[13].

\subsection{ Interface Conditions}
We consider three interface conditions that generate escalation or closure decision of each alert.
\begin{itemize}
    \item C0, Default Threshold (Base Line): 
    
    Decision rule: Escalate in case $p \geq 0.5$ otherwise Close.
None of the calibration, no uncertainty cue.
    \item C1, Misaligned Trust (Over-trust Confidence): Uncalibrated confidence is revealed through the interface [28].

Decision rule: Escalate if $p \geq 0.7$, otherwise Close.
It is a failure mode (conservative analysts become  more conservative about failure as uncertainty increases, unless they are very sure that failure will happen) that is reflected in this model: low confidence costly misses are unacceptable in the actual operational context of the situation, but these high confidence costly misses are tolerated.The 0.7 threshold represents conservative escalation behavior as witnessed in the SOC setting in which analysts hedge only when they consider the confidence to be high.
\item C2, Aligned Trust (Proposed): The interface reveals calibrated confidence with uncertainty category.Also it gives the decision recommendation based on the cost model.

Decision rule: Escalate if $p_{\mathrm{cal}} \geq t^{*}$

Safety override: in case of uncertainty, High, then increase even in the case of $p_{\mathrm{cal}} < t^{*}$

This is the explicit application of decision-conscious alignment and reserved treatment of dubious cases.

\end{itemize}

\subsection{Decision-Aware Cost Model and Threshold}
\FloatBarrier
\begin{figure}[!htbp]
\centering
\begin{subfigure}{0.48\linewidth}
  \centering
\includegraphics[width=0.9\linewidth]{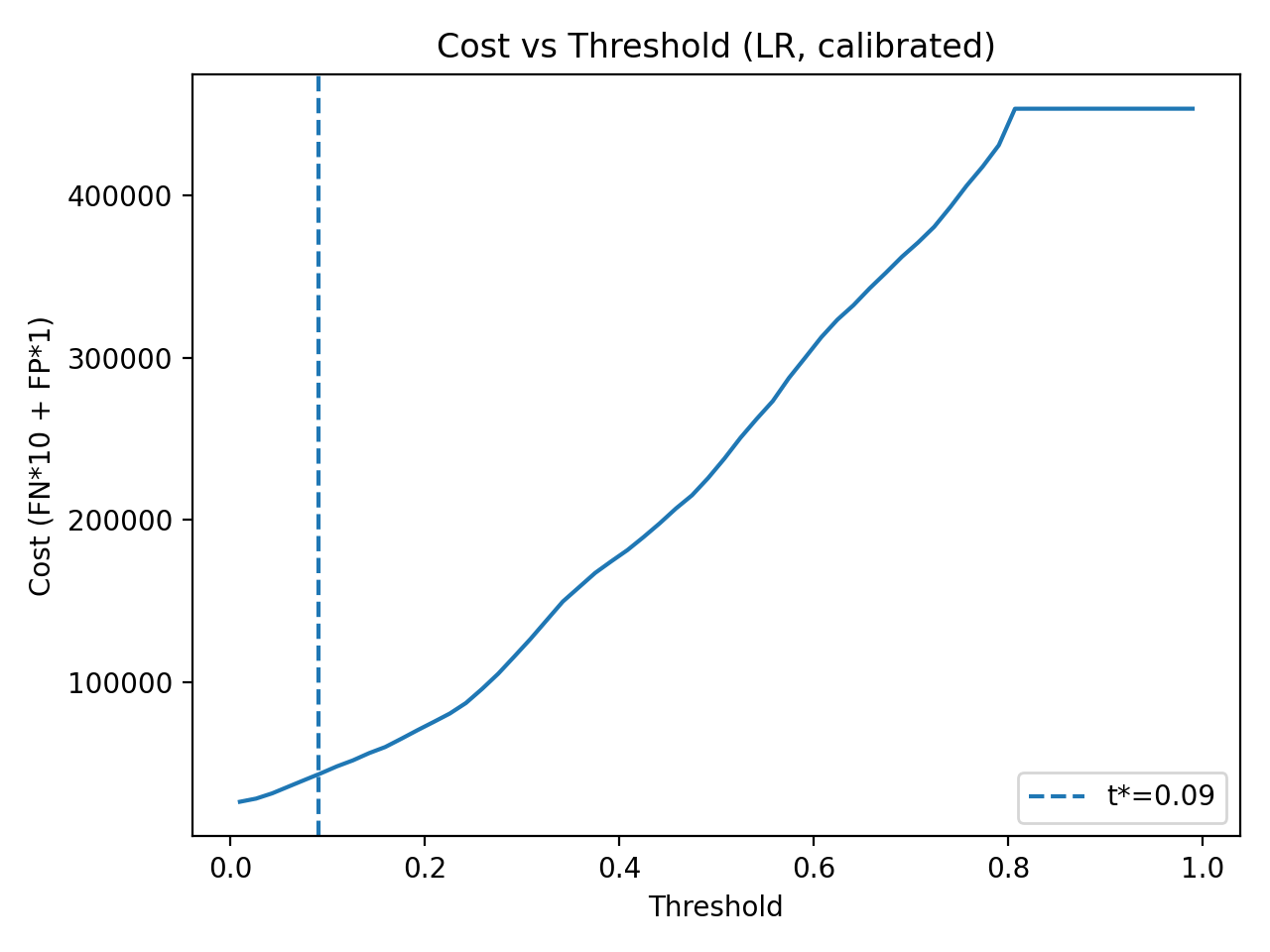}
  \caption{Logistic Regression (LR).}
  \label{fig:cost_lr}
\end{subfigure}
\hfill
\begin{subfigure}{0.48\linewidth}
  \centering
\includegraphics[width=0.9\linewidth]{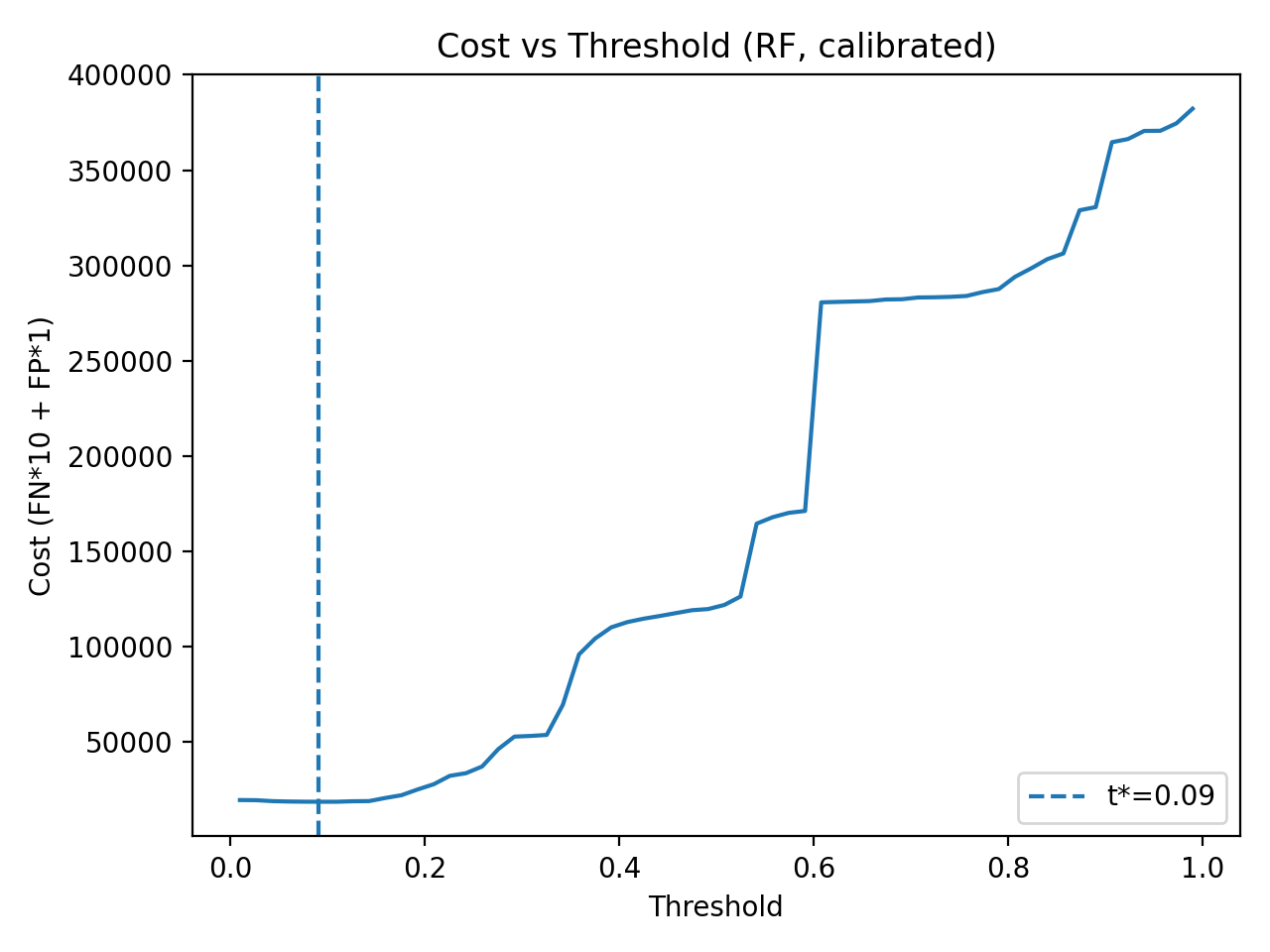}
  \caption{Random Forest (RF).}
  \label{fig:cost_rf}
\end{subfigure}
\caption{Cost-weighted loss as a function of the escalation threshold using calibrated probabilities. The dashed vertical line indicates the decision-aware threshold $t^* = 0.0909$ derived from asymmetric costs ($C_{FN}=10$, $C_{FP}=1$).}
\label{fig:cost_threshold}
\end{figure}
\FloatBarrier

Figure~\ref{fig:cost_threshold} illustrates the resulting cost-weighted loss as a function of the escalation threshold for both Logistic Regression and Random Forest using calibrated probabilities.
In both cases, expected cost increases rapidly as the threshold rises, reflecting the sharp growth in false negatives under conservative escalation policies.

as We set: ($C_{FN}=10$, $C_{FP}=1$); Under this cost model, the decision-aware escalation threshold is given by
\[
t^* = \frac{C_{FP}}{C_{FP} + C_{FN}} = \frac{1}{1 + 10} = 0.0909.
\]
The dashed vertical line in Fig.~\ref{fig:cost_threshold} indicates this threshold, which corresponds closely to the region of minimal expected cost for both models.This is the only threshold in C2 (Aligned).

SOC is an unbalanced triage: false alarms are usually much less serious than missed attacks. We take the explicit cost model that is consistent with the cost sensitive decision theory [11],[15],[23].

\subsection{Simulation-Based Results}
The asymmetric cost ratio ($C_{\mathrm{FN}} \gg C_{\mathrm{FP}}$) is indicative of common SOC practice, in which down-stream cost of missed intrusions is normally significantly greater than that of false alerts. We take a representative ratio to show the impact of decisions but the proposed framework can be extended to cost structures of arbitrary structure and can be adjusted to the risk tolerance of an organization.

Table~\ref{tab:results} presents the simulation-based triage outcomes for all
interface conditions and models. False negatives (FN), false positives (FP), and
the resulting cost-weighted loss ($10\cdot FN + 1\cdot FP$) are reported to
highlight the impact of trust signal alignment on cost-sensitive decision
performance.

\begin{center}
\captionof{table}{SOC triage outcomes under baseline, misaligned, and aligned trust interfaces. Cost = $10\cdot FN + 1\cdot FP$.}
\label{tab:results}
\begin{tabular}{llrrrrrr}
\toprule
Model & Condition & FN & FP & Cost \\
\midrule
LR & C0 Baseline & 23693 & 12959 & 249889 \\
LR & C1 Misaligned & 32490 & 9285 & 334185 \\
LR & C2 Aligned & 2286 & 20396 & 43256 \\
RF & C0 Baseline & 27400 & 12034 & 286034 \\
RF & C1 Misaligned & 27509 & 7681 & 282771 \\
RF & C2 Aligned & 77 & 18007 & 18777 \\
\bottomrule
\end{tabular}
\end{center}

As shown in Table~\ref{tab:results}, the misaligned trust condition (C1)
substantially increases false negatives relative to the baseline, despite
reducing false positives. In contrast, the aligned trust condition (C2)
consistently minimizes cost-weighted loss across both models by aggressively
reducing false negatives.

Our results are calculated by False Negatives (FN), False Positives (FP) and cost-weighted loss:

Cost =10.FN + 1.FP

    Logistic Regression (LR): 

    \begin{itemize}
        \item C0 Baseline: FN = 23,693, FP = 12,959, Cost = 249,889
        \item C1 Misaligned (0.7): FN = 32,490, FP = 9,285, Cost = 334,185
        \item C2 Aligned (.0909 + uncertainty override): FN = 2286 FP = 20396 Cost = 43256
    \end{itemize}

 When the raw confidence is exposed and strict threshold (C1) is applied, the FN value increases significantly compared to the baseline and the decision-aware alignment (C2) makes the FN value significantly decrease and the cost-weighted loss decrease to the lowest value.

 Random Forest (RF):
    
\begin{itemize}
    \item C0 Baseline: FN = 27,400, FP = 12,034, Cost = 286,034
    \item C1 Misaligned (0.7): FN = 27,509, FP = 7,681, Cost = 282,771
    \item C2 Aligned (0.0909 + uncertainty override): FN = 77, FP = 18,007, Cost =18,777
\end{itemize}

\begin{figure}[!t]
\centering
\begin{subfigure}{0.48\linewidth}
    \centering
\includegraphics[width=0.9\linewidth]{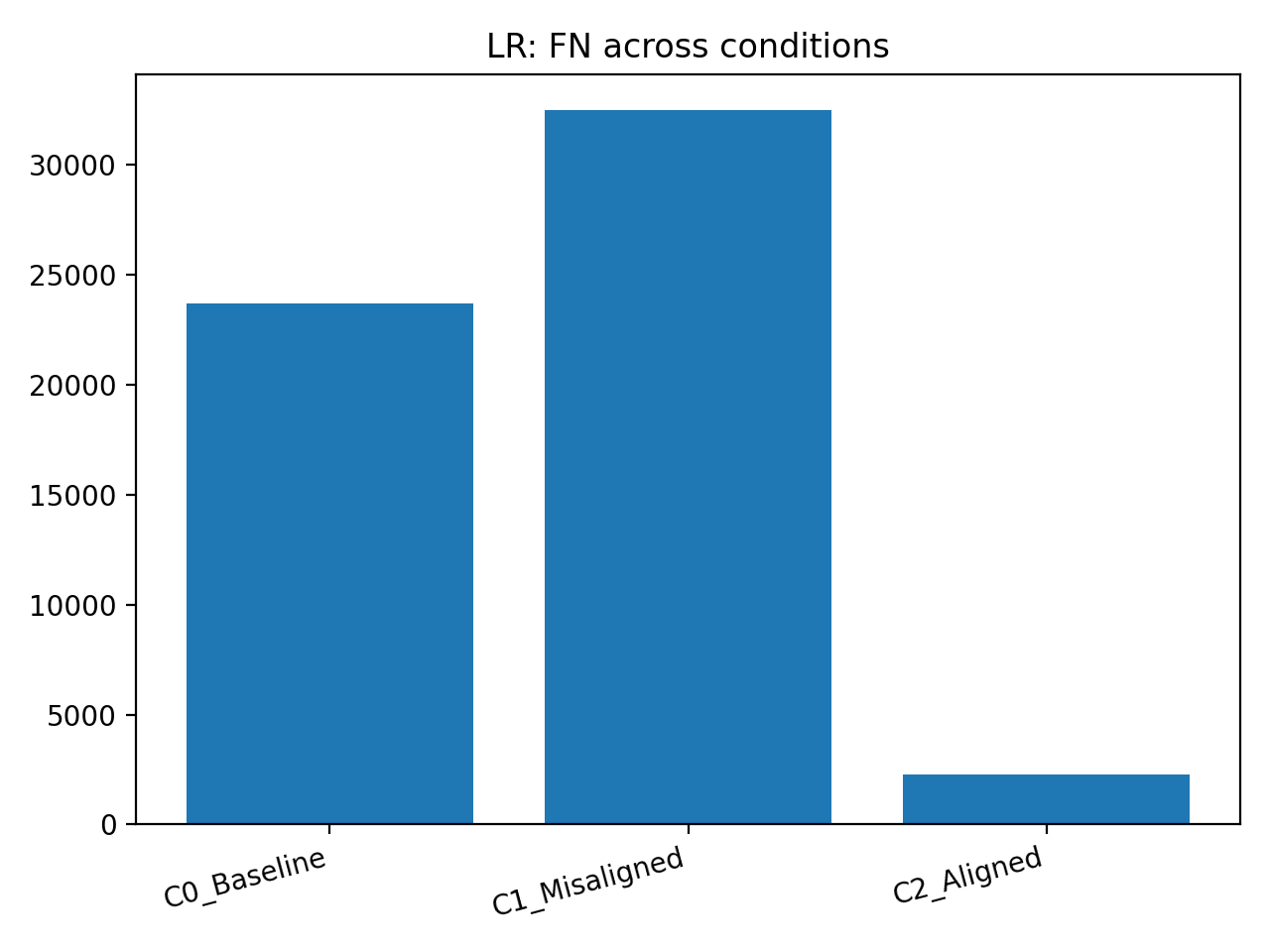}
    \caption{Logistic Regression (LR).}
    \label{fig:lr_fn}
\end{subfigure}
\hfill
\begin{subfigure}{0.48\linewidth}
    \centering
\includegraphics[width=0.9\linewidth]{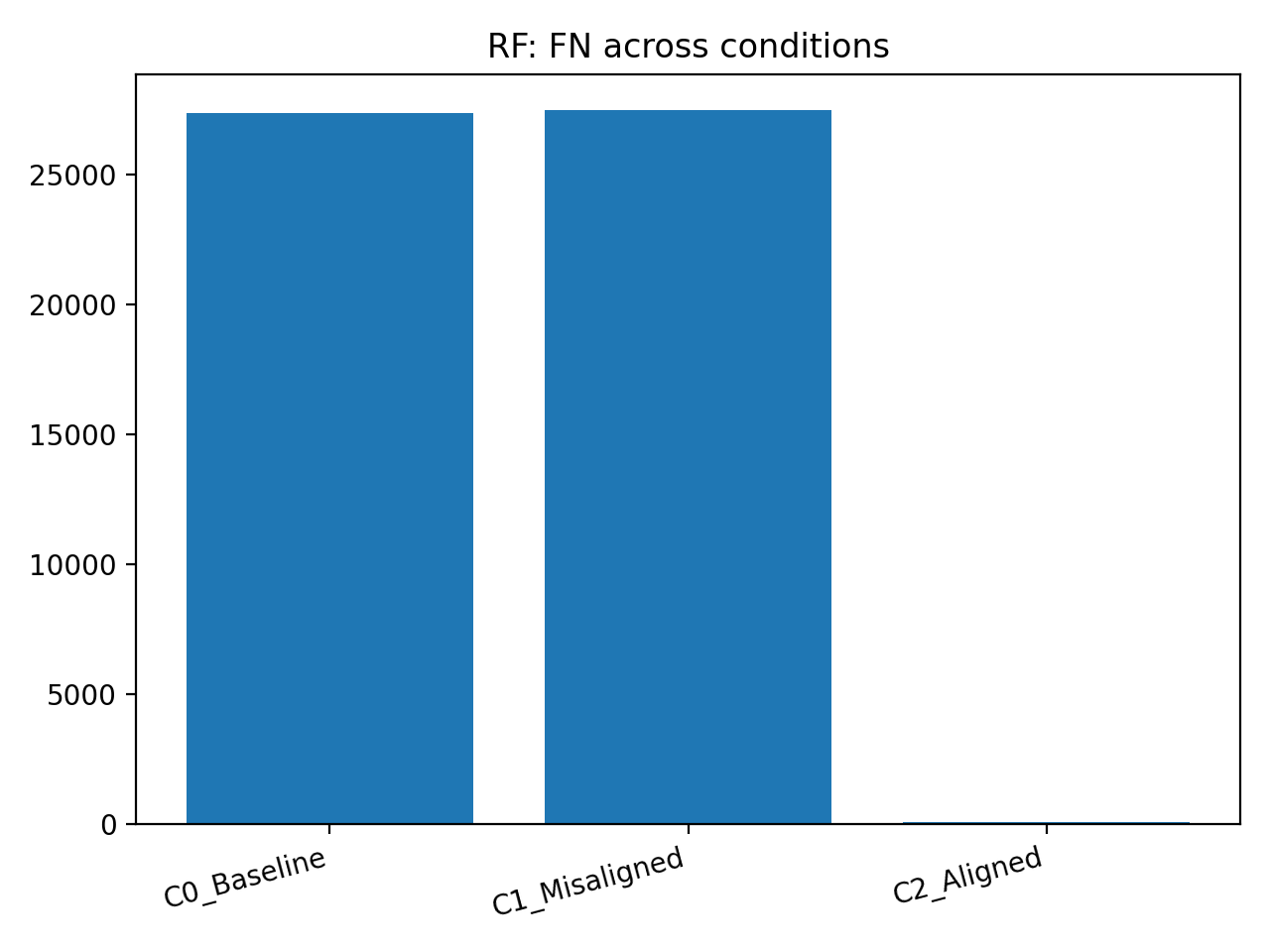}
    \caption{Random Forest (RF).}
    \label{fig:rf_fn}
\end{subfigure}

\label{fig:fn_comparison}
\end{figure}
Figure-4, False negatives (FN) across interface conditions for Logistic Regression and Random Forest.The aligned trust interface (C2) substantially reduces missed attacks compared to baseline (C0) and misaligned (C1) conditions.

We confirmed that the qualitative behavior of decision-aware thresholding is still valid to a variety of asymmetric cost ratios (CFN/CFP between 5:1 and 20:1), where aligned trust interfaces will always optimize expected cost. This indicates a high level of robustness of the proposed alignment mechanism than for just one cost configuration.

The chart visualizes the number of false negatives under the three interface conditions for both detection models.
Across both Logistic Regression and Random Forest, the aligned trust interface (C2) leads to a dramatic reduction in missed attacks compared to the baseline (C0) and misaligned (C1) conditions.
This effect is consistent with the use of calibrated confidence and a decision-aware escalation threshold, which together encourage conservative handling of uncertain alerts.

The same pattern occurs in a different model family, which proves the argument that alignment is a model-agnostic decision-support layer.

\subsection{Human-in-the-Loop Study Protocol}

In order to augment the findings of simulation and assess the hypothesis that decision-conscious trust signal congruence enhances human triage, we introduce a controlled human-in-the-loop protocol capturing two realities of SOC: (i) high alert volume and alert fatigue, and (ii) the operational risk of lost attacks. Past empirical studies have recorded that SOC tooling can cause high false-alarm burdens and that alerts have to be validated by the analysts putting significant effort into it [3],[4]. Another SOC challenge that has continued to be persistent is alert fatigue, which has impacted on the quality of decision and the management of workload [3],[17].

To have a possibility of controlled validation of the effects of a trust alignment, the human-in-the-loop protocol has been annotated; this optimization is planned on future work because of the resource and IRB limitations.
\subsubsection{Participants}
We are taking 40 participants divided into two groups:
\begin{itemize}
\item 25 proxy analysts were recruited based on Ontario Tech University (graduate level cybersecurity or closely related courses).
\item 15 practitioners were hired to an external industry company (security operations, IT security or related positions).
\end{itemize}
\subsubsection{Conditions of Materials and Interfaces}
Trial conditions Triage under three interface conditions are administered to the participants:
\begin{itemize}
    \item C0 (Baseline): bare-faced interface (none of the explicit confidence/uncertainties).
    \item C1 (Misaligned): uncoated raw display of confidence.
    \item C2 (Aligned, Proposed): confidence, uncertainty, cost-mindful decision recommendation, where high uncertainty has conservative escalation.
\end{itemize}
The content and presentation features of every condition are the same, with the only difference being the presentation of the trust-signals.

\subsubsection{Task and Procedure}
In a within-subjects design, each participant is exposed to a set of fixed alerts that he/she has to triage in all three conditions:
\begin{itemize}
    \item Orientation (5-8 minutes): a set of short instructions where the objective of the triage (Escalate vs Close) and the meaning of the elements on the interface are described.
    
        \item Condition blocks: triage of the participants proceeds in a counterbalanced (Latin-square form of rotation) order C0/C1/C2 to overcome the impacts of learning and fatigue across the conditions.
    \end{itemize}
\subsubsection{The behaviors of Per Trial}
 participants choose Escalate or Close.
 \begin{itemize}
     \item optionally give a rating of confidence (e.g. 1-5 Likert scale)
     \item go to the next alert (no reply to whether it is correct or not)
     \item In order to minimize the effect of memory, the alerts are randomly shuffled in each block of condition but with the same set of alerts across conditions.
 \end{itemize}
     \subsubsection{Logging}
     For each trial, we record: 
     \begin{itemize}
    \item participant identifier and participant group (Ontario Tech proxy analysts or industry practitioners),
    \item interface condition (C0, C1, or C2),
    \item alert identifier together with the corresponding ground-truth label,
    \item participant decision to \emph{Escalate} or \emph{Close} the alert,
    \item decision time recorded per alert.
\end{itemize}

      \subsubsection{Primary and Secondary Outcome Measures}
      Cost-sensitive SOC risk and Primary outcomes:
\begin{itemize}
    \item false positives (FP) and false negatives (FN),
    \item cost-weighted loss computed using the same asymmetric cost model applied in the simulation-based evaluation.
\end{itemize}

Secondary outcomes evaluate efficiency as well as subjective trust behavior:
\begin{itemize}
    \item decision time per alert and aggregated per interface condition,
    \item calibration between self-reported confidence and decision correctness, computed by comparing confidence ratings with empirical accuracy.
\end{itemize}

\subsubsection{Analysis Plan}
Because all participants are exposed to all interface conditions, a within-subject (paired) analysis design is employed.
The primary analysis focuses on differences between the baseline condition (C0) and the aligned trust condition (C2).
Non-parametric paired statistical tests are used, as no assumptions are made regarding the distribution of behavioral measures.
Statistical significance is evaluated at standard confidence levels, and results are reported together with descriptive statistics to facilitate interpretation~\cite{wilcoxon1945individual}.

\section{Discussion}
This section is the interpretation of the findings of the experiment and the reasoning as to why trust signal alignment based upon decision awareness increases SOC alert triage performance. Instead of restating the numbers, the discussion is devoted to the mechanisms of failure and success in different interface conditions and draws general conclusions on the design of SOC systems [6],[24].
\subsection{Why Misaligned Trust Fails}
The misaligned trust condition (C1) is worse both in comparison to the baseline condition and also to the aligned condition, which can be attributed in the first place to the amplification of overconfidence and misuse of thresholds. In the case where the raw and uncalibrated confidence scores have been exposed without any contextual clues, users are more likely to perceive high confidence as a good predictor of the correctness of the answer, despite these scores having a poor correlation with the actual risk [6],[22],[26]. There is previous evidence of humans often over-trusting obviously quantified probabilistic output, particularly when pressed in time, resulting in automation bias and decision complacency [14],[19].

The use of excessive severe escalation thresholds also adds to this effect in the context of SOC triage. As was seen in the simulation outcome, with a higher confidence threshold (e.g. 0.5 to 0.7), there are fewer false positives but there are much more false negatives. This action is indicative of a general mental shortcut associated with the identification of a certain level of cognitive confidence and the belief that it corresponds to a certain degree of importance and an overall dismissal of the asymmetry of cost of the missed attacks. This misuse of thresholds ends up enhancing human bias instead of reducing it leading to poor quality of decisions despite the availability of other information.
\subsection{Why Aligned Trust Works}
In comparison, the aligned condition of trust (C2) enhances when the trust signals are directly connected to the cost of decisions instead of directly to the raw model output. Calibrated confidence is used to make probability values closer to empirical likelihoods, reducing systematic overconfidence. Even more to the point, the fact that the cost-sensitive decision threshold recasts confidence as a decision-relative measure is used to guide escalation behavior in a way that is consistent with the operational risk [20],[25].
The explicit uncertainty signal is also added, which also helps to enhance the performance. Kind of alert that are in the proximity of the decision boundary are marked as uncertain and conservatively inflated so that variations of marginal confidence do not lead to early dismissal. The present uncertainty-conscious escalation methodology is consistent with the previous results that demonstrate that cautious treatment of uncertain situations is advantageous in high stakes areas [14]. The combination of cost awareness and uncertainty management causes the decision-making goal to be changed towards minimizing operations harms as opposed to false alarms hence the cost weighted loss and false negative is significantly minimized under the aligned condition.
\subsection{Implications for SOC Design}
These results directly relate to SOC tool and interface design. To start with, they show that user interface design can be such an ingredient as model performance in terms of operational outcome. Although the detection models may be kept constant, changing the presentation or interpretation of the trust signals can result in changes in cost-sensitive performance by orders of magnitude.

Second, the findings indicate that trust cannot be considered as an absolute characteristic of a model, but as a relative one. The value of confidence can only have its meaning when considered with the cost of decision and uncertainty. Interfaces exposing crude confidence out of context dispose of inviting improper dependency or rejection [15]. Lastly, the results show that upgrades in incremental models should not be considered sufficient in solving SOC decision failures [27]. In the absence of a similar advancement in terms of trust alignment and decision support, superior models can still produce inferior human-AI interaction.
\subsection{ Model-Agnostic Insights}
Another remarkable result of the assessment is that the identical qualitative patterns are observed in both of the Logistic Regression and the Random Forest models [3]. Although both models have differences in their representational ability and error distributions, they have higher false negatives in the case of misaligned trust and their costs are minimized in the case of aligned trust. Such consistency indicates that the effects, which are being observed do not depend on one particular classifier, but rather, it is the result of the interaction of the trust signals and the decision logic [14].

These results are model-agnostic, which enhances the applicability of the proposed framework. Since trust alignment is executed as a post-processing and interface-level framework, it can be executed on a diverse variety of detection models without retraining or architectural change. This is very flexible and thus the approach is of special relevance to real-world SOC deployments, where it is frequently not feasible to replace existing models.

\section{Limitations and Future Work}

\subsection{Limitations}

Regardless of the promising outcomes, there are some limitations of this study that need to be admitted.

To begin with, the major assessment is based on the simulated behavior of analysts and not the observations of SOC analysts in practice. Although the analysis by simulation is a prevalent tool to test the policies of making decisions and behavior of the system on the large scale, it cannot adequately reflect the cognitive strategies, situational reasoning, and organizational restrictions that affect the security operation in real-life situation. It has been demonstrated in the previous research that experience, workload, and situational awareness influence the decision-making of analysts and are challenging to reflect in terms of the simulation modeled explicitly [18].

Second, the experimental test is done on one benchmark data (UNSW-NB15). This dataset is generalized in the research of intrusion detection and has a variety of attack types despite being one of the most popular data sets that may not be representative of the changing threat environment and the diversity of enterprise network settings. Absolute performance results can be affected by dataset-specific characteristics like class imbalance, feature distributions, and labeling practices. They are dataset-specific characteristics, which can affect absolute performance results including class imbalance, feature distributions, and labeling practices. The dataset-specific characteristics can be one of the issues that can influence absolute performance results: class imbalance, feature distributions, and labeling practices. Such issues as class imbalance, feature distributions, and labeling practices are dataset-specific characteristics that can affect absolute performance results.

Third, the paper is concerned with a binary choice (escalate vs. close). Though this abstraction matches with typical SOC triage processes, in practice the decision space that real-world alert management processes can use is more detailed, incorporating more aspects such as prioritization, delayed investigation, and partial escalation. Consequently, the gains that have been reported are mainly attributed to the improvement of conservative escalation decisions and not complete incident-response pipelines.

\subsection{Future Work}

These limitations give rise to a number of promising research directions.

One important follow-up measure is human-in-the-loop testing of professional SOC analysts, which enables the direct determination of trust calibration, cognitive workload, and confidence in decisions in congruent and incongruent trust interfaces. The controlled user studies would allow studying the impact of the calibrated confidence and uncertainty signals on the behavior of the analysts in longer working periods in greater detail 

The framework should also be expanded in the future to multi-class and multi-stage alert environments, where the analysts will have to distinguish between the types of attacks, level of their severity and the level of urgency over responding to the attack [23]. A hierarchical or sequential decision process should consider decision-conscious trust alignment to be more realistic of SOC workflows, paths toward incident escalation.

The other direction that is significant is the implementation of changing and context-specific cost models. As a practical matter, the cost of false negatives and false positives can change over time depending on the importance of the asset, threat intelligence or in the context of operational setting. Operational robustness could be further enhanced by adaptive cost models that adjust decision thresholds to varying risk profiles over time, as well as by cost models that are more adaptive to risk profile changes, and that re-establish the original decision threshold in the future upon encountering fresh risks and opportunities [21]

Lastly, in the future, the adaptive trust interfaces could be investigated in order to adjust the display of confidence, uncertainty and explanations according to analyst expertise, workload or past patterns of interaction. These interfaces are potentially useful to avoid long-term over- and under-trust and enable long-term human-AI partnership in security activities 

\section{Conclusion}
Model outputs presented and acted upon by human analysts has emerged as a major issue with machine-learning-based alerting as a core part of contemporary Security Operations Centers (SOCs). Specifically, the use of confidence scores is frequently presented irrespective of the costs associated with decisions or the uncertainty, and as a result, systematic over-trust or under-trust and, subsequently, the use of costly triage errors are made. This is particularly an issue in SOC environments, where false negatives of attacks are highly operational in nature [15],[21].

We have introduced a decision-conscious trust signal alignment framework in this work, which directly relates model certainty, calibration, and uncertainty with cost-sensitive cost escalation decisions. Instead of adjusting the models of detection, the method works on the interface level, matching the trust signals with asymmetric error costs using calibrated probabilities and a decision threshold that is theoretically based. The framework is model-agnostic and does not need retraining or any architectural modifications, as it can be used with the existing SOC pipelines.

With a contrast to the previous studies that focus on the detection accuracy or an independent trust mechanism, this paper restructures trust as a decision-consistent control issue through the connection between model confidence and operational costs and analyst behavior.

The analysis of the UNSW-NB15 data with the help of the Logistic Regression and Random Forest models shows significant changes in the decision results. In each of the models, aligned interface condition (C2) lowers the false negative by more than an order of magnitude over the baseline and misaligned conditions, and also obtains a drastic reduction in cost-weighted loss under the model of $C_{\mathrm{FN}} = 10,\; C_{\mathrm{FP}} = 1$

These profits are obtained without the need to have more complex decisions, demonstrating the usefulness of decision-conscious trust alignment.

Combined, the findings are indicative of the fact that most SOC failures due to model constraints, perhaps atherosclerotic, are in fact due to mismatched trust interface. The explicit inclusion of decision costs in the process of communicating confidence and uncertainty allows SOCs to have safer and more conservative triage behavior without making the model less general or deployable. A possible practical and model-agnostic way to achieve safer human-AI interaction in cybersecurity activity consists in aligning the trust signals with the costs of the decision-making process [19],[22].

This work provides a feasible avenue in the deployment of AI systems that analysts can depend on when facing operational pressure by the simple concept of reframing trust as a decision-congruent signal as opposed to being a model attribute.

\end{document}